\documentstyle[12pt]{ioplppt}
\jl{6}

\def \ep{\varepsilon}
\def \c{\mbox{curl}\,}
\def \d{\mbox{D}}   
\def \D{\mbox{D}}
\def \ts{\textstyle}
\def \rd{\displaystyle{\cdot}}
\def \div{\mbox{div}\,}
\def \i{{\cal I}}

\begin{document}

\title{Gravito-electromagnetism}

\author{Roy Maartens\dag \ftnote{4}{maartens@sms.port.ac.uk}
and Bruce A. Bassett\ddag\S \ftnote{5}{bruce@stardust.sissa.it}
}

\address{\dag\ School of Computer Science and Mathematics, 
Portsmouth University, Portsmouth~PO1~2EG, Britain
}

\address{\ddag\ International School for Advanced Studies, SISSA/ISAS,
Via Beirut~2-4, Trieste~34014, Italy
}

\address{\S\ Department of Maths and Applied Maths, University of 
Cape Town, Rondebosch~7700, South Africa
}

\begin{abstract}

We develop and apply a fully covariant $1+3$ 
electromagnetic analogy for gravity.
The free gravitational field is
covariantly characterized by the Weyl gravito-electric
and gravito-magnetic spatial tensor fields, whose dynamical
equations are the Bianchi identities. Using
a covariant generalization of spatial vector algebra and
calculus to spatial tensor fields, we exhibit the covariant
analogy between the tensor Bianchi equations and the vector
Maxwell equations. We identify gravitational source terms,
couplings and potentials with
and without electromagnetic analogues.
The nonlinear vacuum Bianchi equations are 
shown to be invariant under covariant spatial duality rotation
of the gravito-electric and gravito-magnetic tensor fields.
We construct the super-energy density
and super-Poynting vector of the gravitational field
as natural $U(1)$ group invariants, and derive their
super-energy conservation equation.
A covariant approach to gravito-electric/magnetic
monopoles is also presented.

\end{abstract}

\pacs{04.30.-w, 04.20.-q, 11.30.-j, 11.30.Fs, 11.15.Tk}


\section{Introduction}

There is a surprisingly rich and detailed correspondence between 
electromagnetism and General Relativity, uncovered in a series of
fundamental papers by Bel \cite{bel}, Penrose \cite{p}
and others \cite{mat,pir,tr,eh,haw,cm,ell71} 
(see \cite{ell71,zak} for more
references), and further developed recently (see, e.g., 
\cite{jcb,dk,bonnor,mmq,bs,EH96,he,DBE96,mes}).
This correspondence is reflected in the Maxwell-like form of the
gravitational field tensor (the Weyl tensor), the
super-energy-momentum tensor (the Bel-Robinson tensor) and the
dynamical equations (the Bianchi identities).
Another form of the correspondence arises in
the search for geons (localized, non-singular, 
topological solutions 
of Einstein's field equations with mass and angular momentum):
in the known (approximate) solutions,
the geometry of 
the electromagnetic geon is 
identical to that of the   
gravitational geon \cite{wh,br}.

Here we pursue
the `electromagnetic' properties of 
gravity in areas which have 
already proved useful  for extensions of electromagnetism to 
non-Abelian gauge theories and string theory.
Our emphasis is on a $1+3$ covariant, physically transparent,
and non-perturbative
approach, with the gravito-electric and gravito-magnetic
spatial tensor fields as the fundamental physical variables. 
Using an improved covariant formalism, including a covariant
generalization to spatial tensors of spatial vector algebra
and calculus, we show in detailed and
transparent form the correspondence
between the electric/ magnetic parts of the
gravitational field and of the Maxwell field. We
identify gravitational source terms, couplings and potentials
with and without
electromagnetic analogues, thus providing further physical
insight into the role of the kinematic quantities shear, vorticity 
and four-acceleration.

In the vacuum case, we show that the
nonlinear (non-perturbative) Bianchi equations for the 
gravito-electric and gravito-magnetic fields are invariant under 
covariant spatial duality
rotations, in exact analogy with the source-free Maxwell equations
for the electric and magnetic fields.
The analogy is of course limited by the fact that the Maxwell 
field propagates on a given spacetime, whereas the gravitational
field itself generates the spacetime.
The electromagnetic vectors fully characterize a Maxwell solution,
and duality
maps Maxwell solutions into Maxwell solutions. 
The gravito-electric/magnetic tensors are not sufficient to
characterize covariantly a solution of Einstein's equations --
one needs also the kinematic quantities which are subject to
the Ricci identities \cite{ell71}.
Duality is an invariance only of
the Bianchi identities, and not the Ricci identities, so
that it does not map Einstein solutions into Einstein solutions.
Nevertheless, the covariant gravito-electric/magnetic duality
reveals important properties of the gravitational field.

The covariant $1+3$ duality has not to our knowledge been given
before. Although duality invariance follows implicitly from
Penrose's spinor formalism \cite{p,pr}, this is in terms of the
4-dimensional Weyl spinor, rather than its $1+3$
electric and magnetic tensor parts. Four-dimensional covariant tensor
approaches to the electromagnetic analogy have been developed
(see e.g. \cite{dk}), and non-covariant linearized Maxwell-type
equations are well established, both in terms of 
gravito-electromagnetic 
vectors (see e.g. \cite{bct,z}) and tensors (see e.g. \cite{cm}).
In \cite{ln}, a covariant and
nonlinear vector approach is developed for stationary spacetimes.
Our approach is fully covariant and 
non-perturbative, 
and in addition is
centred on the gravito-electromagnetic
spatial tensor fields, allowing for a more direct and transparent 
interpretation based on the Maxwell vector analogy. This approach
is a development of the work by Tr\" umper \cite{tr},
Hawking \cite{haw} and Ellis \cite{ell71}, and is related to
recent work on a covariant approach to gravitational waves
\cite{EH96,he,DBE96,mes} and to local freedom in the
gravitational field \cite{mes}.
A shadow of our general duality result
arises in 
linearized  gravitational  wave theory, where for vacuum or de Sitter 
spacetime, there is an interchange symmetry between the 
gravito-electric and 
-magnetic tensors \cite{DBE96}.

Duality invariance has important implications in field theory 
in general.
It was essentially this symmetry of the Abelian theory, and attempts 
to 
extend it to include matter, which led to the Montonen-Olive 
electromagnetic duality conjecture that there exists a group 
transformation
mapping electric monopoles into 
magnetic monopoles within the framework 
of a non-Abelian (specifically $SU(2)$) 
gauge theory \cite{MO}. This 
conjecture has proved particularly fruitful, stimulating work 
on  $S$, $T$ 
and $U$ dualities in string 
theory  (see e.g. \cite{pol,sen,vaf}), the 
extension of the    electromagnetic
duality to magnetically charged black holes and nonlinear
electrodynamics \cite{deser,gr} and leading to the
Seiberg-Witten proof of quark confinement in
supersymmetric Yang-Mills theory via monopole
condensation \cite{SW}.

We use the covariant spatial duality to
find the gravitational super-energy density and 
super-Poynting vector as natural group invariants, and
derive a new covariant super-energy conservation equation.
Finally, we discuss gravito-electric/magnetic
monopoles, providing a 
covariant characterization,
in contrast to previous non-covariant treatments \cite{dn,dr,z}. 
In the linearized case, we
show that  the Taub-NUT gravito-magnetic monopole given in
\cite{dn} is related to
the Schwarzschild gravito-electric monopole 
by a spatial duality rotation and  an interchange of four-acceleration 
and vorticity. This provides a covariant form of the
relation previously given in non-covariant 
approaches \cite{ger,dowker}. It is well-known that the NUT metrics
may be obtained from the Schwarzschild metric via the Ehlers-Geroch
transformation \cite{ger}. This transformation is 
in fact the generator of $T$-duality in string theory,
but it is not a duality transformation
in the sense described here, since it maps Einstein solutions to
Einstein solutions and thus necessarily involves 
kinematic and geometric conditions in addition to a duality
rotation. Furthermore, the Ehlers-Geroch transformation requires
the existence of a Killing vector field, whereas the general
duality that we present does not require any spacetime
symmetry.

\section{Covariant spatial vector and tensor calculus}

To elaborate the electromagnetic
properties of the free gravitational field in General Relativity, 
we first 
present the required covariant formalism, which is based on 
\cite{roy}, a streamlined and extended version of the Ehlers-Ellis
$1+3$ formalism \cite{ell71}.
Then we give the covariant
form of the Maxwell spatial duality in a general
curved spacetime. In the following section we extend the
treatment to the gravitational field.

Given a congruence of observers with four-velocity
field $u^a$, then
$h_{ab}=g_{ab}+u_au_b$ projects into the local rest spaces,
where $g_{ab}$ is the spacetime metric.\footnote{We follow
the notation and conventions of \cite{ell71,roy}. 
(Square) round brackets enclosing indices denote (anti-) 
symmetrization, while angled brackets denote the spatially projected, 
symmetric and tracefree part;
$a,b,\cdots$ are spacetime indices.}
The spatially projected part of a vector is 
\[
V_{\langle a\rangle}
=h_a{}^bV_b\,, 
\]
and
the spatially projected, symmetric and tracefree part of a 
rank-2 tensor is 
\[
A_{\langle ab\rangle}=h_{(a}{}^ch_{b)}{}^dA_{cd} 
-{\ts{1\over3}}h_{cd}A^{cd}h_{ab} \,.
\]
The spatial alternating tensor is
\[
\ep_{abc} = \eta_{abcd}u^d=\ep_{[abc]} \,, 
\]
where $\eta_{abcd}=\eta_{[abcd]}$
is the spacetime alternating tensor. Any spatial rank-2 tensor 
has the covariant irreducible decomposition: 
\[
A_{ab}={\ts{1\over3}}h_{cd}A^{cd}h_{ab}+
A_{\langle ab\rangle}
+\ep_{abc}A^c \,,
\]
where  
\[
A_a={\ts{1\over2}}\ep_{abc}A^{[bc]}
\]
is the vector that is the spatial dual to the skew part.
Thus the skew part of a spatial tensor
is vectorial, and the irreducibly tensor part is symmetric.
In the $1+3$ covariant approach \cite{ell71,mes}, 
all physical and geometric
variables split into scalars, spatial vectors or spatial
tensors that satisfy
$A_{ab}=A_{\langle ab\rangle}$. From now on, all rank-2
spatial tensors will be assumed to satisfy this condition.

The covariant spatial vector product is
\[
[V,W]_a=\ep_{abc}V^bW^c \,,
\]
and the covariant generalization to spatial tensors is
\[
[A,B]_a=\ep_{abc}A^b{}_dB^{cd} \,,
\]
which is the vector that is spatially dual to the covariant
tensor commutator.

The covariant time derivative
is 
\[
\dot{A}^{a\cdots}{}{}_{b\cdots}=
u^c\nabla_cA^{a\cdots}{}{}_{b\cdots} \,,
\]
and the covariant spatial derivative is
\[
\D_a A^{b\cdots}{}{}_{c\cdots}=h_a{}^ph^b{}_q\cdots h_c{}^r\cdots
\nabla_p A^{q\cdots}{}{}_{r\cdots} \,.
\]
Then the covariant spatial divergence and curl of vectors
and rank-2 tensors 
are defined by \cite{roy,mes}:
\begin{eqnarray}
&& \div V=\D^aV_a\,,~~~~\c V_a=\ep_{abc}\D^bV^c\,,
\label{dc0} \\
&&(\div A)_a=\D^b A_{ab}\,, ~~~~ \c A_{ab}=
\ep_{cd(a}\D^c A_{b)}{}^d \,,
\label{dc}
\end{eqnarray}
where $\c A_{ab}$ is tracefree if $A_{ab}=
A_{(ab)}$. The tensor curl and divergence are related by
\[
\ep_{abc}\D^bA_d{}^c=\c A_{ad}+{\ts{1\over2}}\ep_{adc}\D_bA^{bc}\,.
\]
The kinematics of the 
$u^a$-congruence are described by the expansion
$\Theta=\D^au_a$, the shear $\sigma_{ab}=\D_{\langle a}u_{b\rangle}$,
the vorticity $\omega_a=-{\ts{1\over2}}\c u_a$, and the 
four-acceleration $\dot{u}_a=\dot{u}_{\langle a\rangle}$.

The above operators obey the covariant identities
\begin{eqnarray}
\left(\D_a f\right)^{\rd} &=& \D_a\dot{f}-{\ts{1\over3}}\Theta
\D_af+\dot{u}_a\dot{f}-\sigma_a{}^b\D_bf-[\omega,\D f]_a
+u_a\dot{u}^b\D_bf\,,
\label{id0}\\
\c \D_af &=& -2\dot{f}\omega_a \,, \label{id1}\\
\D^a[V,W]_a &=& W^a\c V_a-V^a\c W_a \,, \label{id2}\\
\D^a[A,B]_a &=& B^{ab}\c A_{ab}-A^{ab}\c B_{ab} \,, \label{id3}
\end{eqnarray}
together with far more complicated identities \cite{roy,mes}. 
In the case where spacetime is almost spatially
isotropic and homogeneous, i.e. a linearized perturbation of
a Friedmann-Lemaitre-Robertson-Walker (FLRW)
background, some of the main further identities take
the linearized form \cite{eb,mt}
\begin{eqnarray}
\left(\D^aV_a\right)^{\rd} &\approx & 
\D^a\dot{V}_a-H
\D^aV_a \,,\label{id4}\\
\left(\D^bA_{ab}\right)^{\rd} &\approx & \D^b\dot{A}_{ab}
-H \D^bA_{ab} \,,\label{id5} \\
\left(\c V_a\right)^{\rd} &\approx & \c\dot{V}_a-H
\c V_a \,,\label{id6} \\
\left(\c A_{ab}\right)^{\rd} &\approx & \c\dot{A}_{ab}-H
\c A_{ab} \,,\label{id7} \\
\D^a\c V_a &\approx & 0 \,,\label{id8} \\
\D^b\c A_{ab} &\approx & {\ts{1\over2}}\c\left(\D^bA_{ab}\right) \,, 
\label{id9} \\
\c\c V_a &\approx & -\D^2V_a+\D_a\left(\D^b V_b\right)+
{\ts{2\over3}}\left(\rho-3H^2\right)V_a \,, \label{id10} \\
\c\c A_{ab} &\approx & -\D^2A_{ab}+{\ts{3\over2}}\D_{\langle a}\D^c 
A_{b\rangle c}+
\left(\rho-3H^2\right)A_{ab} \,, \label{id11}
\end{eqnarray}
where $H$ 
is the background Hubble rate, 
$\rho$ is the background energy density 
and $\D^2=\D^a\D_a$
is the covariant Laplacian.

The electric and magnetic fields measured by $u^a$ observers 
are defined
via the Maxwell tensor $F_{ab}$ by 
\begin{equation}
E_a=F_{ab}u^b=E_{\langle a\rangle}\,,
~H_a={\ts{1\over2}}\ep_{abc}F^{bc}\equiv
{}^*\!F_{ab}u^b=H_{\langle a\rangle}\,,
\label{max0}\end{equation}
where * denotes the dual. These spatial physically measurable
vectors are equivalent to the spacetime Maxwell tensor, since
\begin{equation}
F_{ab}=2u_{[a}E_{b]}+\ep_{abc}H^c \,.
\label{max}\end{equation}

Maxwell's
equations $\nabla_{[a}F_{bc]}=0$ and
$\nabla^bF_{ab}=J_a$ are given in $1+3$ covariant
form for $E_a$ and $H_a$
by Ellis \cite{ell73}. 
In the streamlined 
formalism, these equations take the simplified form
\begin{eqnarray}
\D^aE_a &=& -2\omega^aH_a+\varrho\,,\label{mdiv}\\
\D^aH_a &=& 2\omega^aE_a\,,
\label{mdiv2}\\
\dot{E}_{\langle a\rangle}-\c H_a &=& -{\ts{2\over3}}\Theta E_a
+\sigma_{ab}E^b   
 -[\omega,E]_a+[\dot{u},H]_a -j_a\,, 
\label{edot}\\
\dot{H}_{\langle a\rangle}+\c E_a &=& -{\ts{2\over3}}\Theta H_a
+\sigma_{ab}H^b     
-[\omega,H]_a-[\dot{u},E]_a\,, \label{hdot}
\end{eqnarray}
where $\varrho=-J_au^a$ is the electric charge density and 
$j_a = J_{\langle a\rangle}$ is the electric current.
In flat spacetime, relative to an inertial congruence ($\Theta=
\dot{u}_a=\omega_a=\sigma_{ab}=0$), these equations take their
familiar non-covariant form.

Introducing the complex electromagnetic spatial vector field
$\i_a=E_a+iH_a$, we see that
in the source-free case ($J_a=0$) Maxwell's equations become
\begin{eqnarray}
\D^a\i_a&=&2i\,\omega^a\i_a\,, \label{mi0}\\
\dot{\i}_{\langle a\rangle}
+i\,\c \i_a&=&-{\ts{2\over3}}\Theta\i_a+\sigma_{ab}\i^b
-[\omega,\i]_a-i\,[\dot{u},\i]_a\,.
\label{mi}\end{eqnarray}
It follows that the source-free Maxwell equations in an arbitrary
curved spacetime, relative to an arbitrary congruence of observers,
are invariant under the covariant global
spatial duality rotation $\i_a~\rightarrow~e^{i\phi}\i_a$, where 
$\phi$ is constant.
The energy density and Poynting vector
\begin{eqnarray}
U&=&{\ts{1\over2}}\i^a\overline{\i}_a
={\ts{1\over2}}(E_aE^a+H_aH^a)\,, 
\label{ue}\\
P_a&=&{1\over 2i}[\overline{\i},\i]_a=[E,H]_a\,,
\label{pe}
\end{eqnarray}
are natural
group invariants. Their invariance also follows 
from the duality invariance of the
energy-momentum tensor 
\cite{pr,ell73}
\begin{equation}
M_a{}^b={\ts{1\over2}}\left(F_{ac}F^{bc}+\,
^*\!F_{ac}\,^*\!F^{bc}\right)\,,
\label{me}\end{equation}
since $U=M_{ab}u^au^b$ and $P_a=-M_{\langle a\rangle b}u^b$.
Using the identity (\ref{id2}), and the
propagation equations (\ref{edot}) and (\ref{hdot}), we find a
covariant energy conservation equation:
\begin{eqnarray}
\dot U+\D^aP_a&=&-{\ts{4\over3}}\Theta U-2\dot{u}^aP_a 
 +\sigma_{ab}\left(E^aE^b+H^aH^b\right)\,.
\label{ec}\end{eqnarray}
This reduces in flat spacetime for inertial observers to the
well-known form $\partial_tU+\div\vec{P}=0$.

A further natural group invariant is
\begin{equation}
\pi_{ab}=-\i_{\langle a}\overline{\i}_{b\rangle}=-E_{\langle a}
E_{b\rangle}-H_{\langle a}H_{b\rangle} \,,
\label{pi1}\end{equation}
which is just the anisotropic electromagnetic pressure \cite{ell73}.
It occurs in the last term of the conservation equation (\ref{ec}),
i.e. $-\sigma_{ab}\pi^{ab}$.

For later comparison with the gravitational case, we conclude this
section by considering the propagation of source-free
electromagnetic waves
on an FLRW background, assuming that $E_a=0=H_a$ in the
background. We linearize and take the curl of equation (\ref{edot}), 
evaluating $\c\c H_a$ by the identity (\ref{id10}) and 
equation (\ref{mdiv2}). 
We eliminate $\c\dot{E}_a$ by
linearizing equation (\ref{hdot}), taking its time derivative, and 
using identity (\ref{id6}). The result is the wave equation
\begin{equation}
\Box^2H_a\equiv-\ddot{H}_a+\D^2H_a \approx
5H\dot{H}_a+\left(2H^2+{\ts{1\over3}}\rho-p\right)H_a \,,
\label{wave1}\end{equation}
where $p$ is the background pressure, and we used the FLRW
field equation $3\dot{H}=-3H^2-{1\over2}(\rho+3p)$. A similar
wave equation may be derived for $E_a$.

\section{The Bianchi identities and nonlinear duality}

The Maxwell analogy in General Relativity is based on the
the correspondence $C_{abcd}\,\leftrightarrow\,F_{ab}$, where
the Weyl tensor $C_{abcd}$ is the free gravitational 
field (see \cite{mes}). For a given $u^a$, it splits irreducibly
and covariantly into
\begin{equation}
E_{ab}= C_{acbd} u^c u^d=E_{\langle ab\rangle}\,, 
~H_{ab} = {}^*\! C_{acbd} u^c u^d =H_{\langle ab\rangle}\,,
\label{eh}
\end{equation}
which are called its `electric' and `magnetic' parts by analogy with 
the Maxwell decomposition (\ref{max0}). These 
gravito-electric/magnetic 
spatial tensors are in principle physically
measurable in the frames of comoving observers, and together
they are equivalent to the spacetime Weyl tensor, since \cite{roy}
\begin{equation}
C_{ab}{}{}^{cd}= 4\left\{u_{[a}u^{[c}+h_{[a}{}^{[c}\right\}E_{b]}{}
^{d]} 
+2\ep_{abe}u^{[c}H^{d]e}+2u_{[a}H_{b]e}\ep^{cde}\,.
\label{weyl}\end{equation}
This is the gravito-electromagnetic version of the expression
(\ref{max}).
The electromagnetic interpretation of $E_{ab}$ and $H_{ab}$
is reinforced by the fact that these fields covariantly
(and gauge-invariantly)
describe gravitational waves on an FLRW background (including the
special case of a flat vacuum background) \cite{haw,eb}.

In the $1+3$ covariant approach to General Relativity \cite{ell71}, 
the fundamental quantities are not the metric 
(which in itself does not provide a 
covariant description), but the kinematic quantities of the
fluid, its energy density $\rho$ and pressure $p$, and the
gravito-electric/magnetic tensors. 
The fundamental equations governing these quantities are
the Bianchi identities and the Ricci identities for $u^a$, with
Einstein's equations incorporated
via the algebraic
definition of the Ricci tensor $R_{ab}$
in terms of the energy-momentum
tensor $T_{ab}$.
We assume
that the source of the gravitational field
is a perfect fluid (the generalization to
imperfect fluids is straightforward).
The Bianchi identities are
\begin{equation}
\nabla^dC_{abcd} = \nabla_{[a}\left(-R_{b]c} + {\ts{\frac{1}{6}}} 
Rg_{b]c}\right) \,,
\label{eq:bianchi}
\end{equation}
where $R=R_a{}^a$ and $R_{ab}=T_{ab}-{\ts{1\over2}}T_c{}^cg_{ab}$. 
The contraction 
of (\ref{eq:bianchi}) implies the conservation equations.
The tracefree part of (\ref{eq:bianchi}) gives
the gravitational equivalents of the Maxwell equations
(\ref{mdiv})--(\ref{hdot}), via
a covariant $1+3$ decomposition
\cite{tr,ell71}. 
In our notation, these take the simplified form:
\begin{eqnarray}
\d^b E_{ab} &=& - 3\omega^bH_{ab}+{\ts{1\over3}}\D_a\rho
+[\sigma,H]_a\,,
\label{eq:dive}\\
\d^b H_{ab} &=&  3\omega^bE_{ab} + (\rho+p)\omega_a-[\sigma,E]_a\,,
\label{eq:divh}\\
\dot{E}_{\langle ab\rangle} - \c H_{ab} &=&
- \Theta E_{ab} + 3\sigma_{c\langle a}E_{b\rangle}{}^c 
-\omega^c\ep_{cd(a}E_{b)}{}^d \nonumber \\  
&&+2\dot{u}^c\ep_{cd(a}H_{b)}{}^d-{\ts{1\over2}}
(\rho+p)\sigma_{ab} \,, \label{eq:edot}\\
\dot{H}_{\langle ab\rangle} +\c E_{ab}&=& 
- \Theta H_{ab} + 3\sigma_{c\langle a}H_{b\rangle}{}^c 
-\omega^c\ep_{cd(a}H_{b)}{}^d \nonumber \\  
&&-2\dot{u}^c\ep_{cd(a}E_{b)}{}^d \,.
\label{eq:hdot}
\end{eqnarray}
These are the fully nonlinear equations in covariant form,
and the analogy with the Maxwell
equations (\ref{mdiv})--(\ref{hdot}) is made strikingly
apparent in our formalism.

Vorticity couples to the fields to produce source terms  in
both cases, but gravity has additional sources from a 
{\em tensor coupling
of the shear} to the field. The analogue of the charge density
$\varrho$ as a source for the electric field, is the energy
density spatial gradient $\D_a\rho$ as a source for
the gravito-electric field. 
Since $\D_a\rho$ covariantly describes
{\em inhomogeneity} in the fluid,
this is consistent with the
fact that the gravito-electric field is the generalization of
the Newtonian tidal tensor \cite{ell71}.

There is no magnetic charge source for $H_a$, but 
the gravito-magnetic field $H_{ab}$ has the
source $(\rho+p)\omega_a$.
Since $\rho+p$ is the relativistic 
inertial mass-energy density \cite{ell71}, 
$(\rho+p)\omega_a$ is the
{\em `angular momentum density'}, which we identify as
a gravito-magnetic `charge' density.
Note however
that angular momentum density does not always generate a
gravito-magnetic field. The
G\"odel solution \cite{ell71}
provides a counter-example, where $H_{ab}=0$ and the non-zero
angular momentum
density is exactly
balanced by the vorticity/ gravito-electric coupling
in equation (\ref{eq:divh}), with $\sigma_{ab}=0$.

For both electromagnetism and gravity, 
the propagation of the fields is determined by
the spatial curls, together with a coupling of 
the expansion, shear, vorticity,
and acceleration to the fields.
The analogue of the electric
current $j_a$ is the gravito-electric `current'
$(\rho+p)\sigma_{ab}$, which is the {\em `density of 
the rate-of-distortion energy'}
of the fluid. There is no magnetic current in either case.

If the Maxwell field is source-free, i.e. $\varrho=0=j_a$, and the
gravitational field is source-free, i.e. $\rho=0=p$, then the
similarity of the two sets of equations is even more apparent,
and only the tensor shear coupling in the case of gravity 
lacks a direct electromagnetic analogue.
(Note that these shear coupling terms govern the
possibility of simultaneous diagonalization of the shear
and $E_{ab}$, $H_{ab}$ in tetrad formulations of general 
relativity \cite{barnes,mle}.)

To obtain the gravitational
analogue of the complex equations (\ref{mi0}) and (\ref{mi}),
which lead to the Maxwell duality invariance, we consider the vacuum
case $\rho =0= p$. In general, $u^a$ is no longer uniquely
defined in vacuum, although
in particular cases (such as stationary spacetimes), there may
be a physically unique choice. However, our results
hold for an {\em arbitrary} covariant choice of $u^a$, 
without any special conditions on the congruence.
By analogy with the complex electromagnetic 
spatial vector $\i_a$, we define
the complex gravito-electromagnetic spatial tensor 
\begin{equation}
\i_{ab} = E_{ab} + i\, H_{ab} \,.
\label{eq:i}
\end{equation}
Then equations (\ref{eq:dive})--(\ref{eq:hdot}) 
reduce to: 
\begin{eqnarray}
\D^b\i_{ab} &=& 3i\,\omega^b\i_{ab}-i\,[\sigma,\i]_a \,,
\label{complex1}\\
\dot{\i}_{\langle ab\rangle}  + i\,\c \i_{ab} &=& 
- \Theta \i_{ab} +3\sigma_{c\langle a}\i_{b\rangle}{}^c \nonumber\\
&&-\omega^c\ep_{cd(a}\i_{b)}{}^d-2i\,\dot{u}^c\ep_{cd(a}\i_{b)}{}^d\,.         
\label{eq:complex2}
\end{eqnarray}
Apart from the increased economy, 
the system is now clearly seen to be invariant 
under the global $U(1)$ transformation:
\begin{equation}
\i_{ab} ~\rightarrow~ e^{i\phi} \i_{ab}\,,
\label{eq:u1}
\end{equation}
which is precisely the tensor (spin-2)
version of the vector symmetry of the source-free
Maxwell equations. We have thus established the 
existence of the covariant spatial duality at the level of
the physically relevant gravito-electric/magnetic fields,
in the general
(non-perturbative, arbitrary observer congruence) vacuum case.
(As with electromagnetism, duality invariance breaks down in
the presence of sources.)

A covariant super-energy density and 
super-Poynting vector arise naturally as
invariants under spatial duality rotation, in direct analogy 
with the Maxwell invariants of equations (\ref{ue}) and (\ref{pe}):
\begin{eqnarray}
U&=&{\ts{1\over2}}\i^{ab} \overline{\i}_{ab} = 
{\ts{1\over2}}\left(E_{ab} E^{ab} + H_{ab} H^{ab}\right)\,,
\label{eq:energy}\\
P_a&=&{1\over 2i}[\overline{\i}, \i]_a = [ E, H]_a\equiv
 \ep_{abc}E^b{}_dH^{cd}\,.
\label{eq:poyn}
\end{eqnarray}
This reflects the duality invariance of the
Bel-Robinson tensor \cite{bel}
\begin{equation}
M_{ab}{}{}^{cd}={\ts{1\over2}}\left(C_{aebf}C^{cedf}+\,
^*\!C_{aebf}\, ^*\!C^{cedf}\right)\,,
\label{mg}\end{equation}
which is the natural covariant definition of the 
super-energy-momentum tensor for the free gravitational field,
since \cite{bel,zak}
\begin{eqnarray}
U&=&M_{abcd}u^au^bu^cu^d\,,\label{ug}\\
P_a&=&-M_{\langle a\rangle bcd}u^bu^cu^d\,. \label{pg}
\end{eqnarray}
The agreement between equations (\ref{ug}) and (\ref{eq:energy})
follows obviously from equation (\ref{mg}) on using 
equation (\ref{eh}). 
However,
it is not obvious that equation (\ref{pg}) agrees with our equation
(\ref{eq:poyn}) for the super-Poynting vector, and one requires
the identity (\ref{weyl}) to show the agreement.

Our expression (\ref{eq:energy}) for the gravitational
super-energy density gives a direct and clear analogy
with the electromagnetic energy density (\ref{ue}). Our expression
(\ref{eq:poyn})
for the gravitational
super-Poynting vector, in terms of the
tensor generalization of the vector product, provides a clearer
analogy with the electromagnetic Poynting vector (\ref{pe}).
The analogy is reinforced
by the fact that $U$ and $P_a$ obey 
a super-energy conservation equation
which is the tensor version of the electromagnetic energy
conservation equation (\ref{ec}). 
To show this, we need the new
covariant identity (\ref{id3}).
Using this and the Bianchi propagation
equations (\ref{eq:edot}) and (\ref{eq:hdot}), we find that
\begin{eqnarray}
\dot U+\D^aP_a &=& -2\Theta U-4\dot{u}^aP_a  
+3\sigma^c{}_{\langle a}\left[E_{b\rangle c}E^{ab}+H_{b\rangle c}
H^{ab}\right] \,.
\label{ecg}\end{eqnarray}
This is the non-perturbative and covariant
generalization of Bel's linearized
conservation equation \cite{bel,zak}: 
$\partial_t U=-\div\vec{P}$.

The last term in the conservation equation (\ref{ecg}) contains
another natural group invariant
\begin{equation}
\pi_{ab}=-\i_{c\langle a}\overline{\i}_{b\rangle}{}^c 
=-E_{c\langle a}E_{b\rangle}{}^c-H_{c\langle a}H_{b\rangle}{}^c  \,,
\label{pi2}\end{equation}
which we interpret as the anisotropic super-pressure
of the gravito-electromagnetic field.

A further covariant
quantity that may be naturally constructed from equation (\ref{eq:i})
is
\begin{eqnarray}
\i_{ab}\i^{ab} &=& (E_{ab} E^{ab} - H_{ab} H^{ab})+2 i\,E_{ab} H^{ab}
\nonumber\\
&=& {\ts{1\over8}}(C_{abcd} C^{abcd} + i C_{abcd}{}^*\!C^{abcd})
\label{eq:isquare}
\end{eqnarray}
which is {\em not} invariant under equation (\ref{eq:u1}).
However it proves very useful in categorizing spacetimes
\cite{kramer} and vanishes in
Petrov type-III and type-N spacetimes \cite{bonnor}
(supporting the existence of gravitational waves in these spacetimes,
since the analogous quantities vanish for purely radiative
electromagnetic fields). Further, the electromagnetic
analogue of the real part of equation (\ref{eq:isquare}),
namely $E_{a} E^{a} - H_{a} H^{a} = - {\ts{\frac{1}{2} }}
F_{ab} F^{ab}$, is just the
Lagrangian density. The analogue of the imaginary part is
$E_{a} H^{a} = \frac{1}{4} F_{ab} {}^*\!F^{ab}$ whose
integral in  non-Abelian gauge theories is proportional to the
topological instanton number.

As pointed out in the introduction, duality rotations preserve
the Bianchi identities in vacuum, 
but not the Ricci identities for $u^a$. This is clearly
apparent from the spatial tensor parts of the Ricci identities
\cite{ell71}, which in our formalism have the simplified form
\begin{eqnarray}
E_{ab} &=& \D_{\langle a}\dot{u}_{b\rangle}-
\dot{\sigma}_{\langle ab\rangle}
-{\ts{2\over3}}\Theta\sigma_{ab}-\sigma_{c\langle a}\sigma_{b\rangle}
{}^c-\omega_{\langle a}\omega_{b\rangle}+\dot{u}_{\langle a}
\dot{u}_{b\rangle} \,, \label{r1} \\
H_{ab} &=& \c\sigma_{ab}+\D_{\langle a}\omega_{b\rangle}+
2\dot{u}_{\langle a}\omega_{b\rangle} \,. \label{r2}
\end{eqnarray}
In order to preserve the Ricci identities, and map 
Einstein solutions to Einstein solutions, one needs to perform
kinematic transformations in addition to the duality rotation. 
An example is presented in the following section.

The electromagnetic analogy suggests a further interesting
interpretation of the kinematic quantities arising from the
Ricci equations (\ref{r1}) and (\ref{r2}).\footnote{Note that
these Ricci equations have the same form in the non-vacuum case.}
In flat spacetime, relative to inertial observers, the electric
and magnetic vectors may be written as
\[
\vec{E}=\vec{\nabla} V -\partial_t \vec{\alpha}\,,~
\vec{H}=\c \vec{\alpha}\,,
\]
where $V$ is the electric scalar
potential and $\vec{\alpha}$ is the
magnetic vector potential.\footnote{The 
covariant form of these potentials is $V=u^aA_a$, 
$\alpha_a=A_{\langle a\rangle}$, where $A_a$ is the four-potential.}
Comparing now with the
 Ricci equations (\ref{r1}) and (\ref{r2}), we see that
the four-acceleration is a covariant
gravito-electric vector potential and the
shear is a covariant gravito-magnetic 
tensor potential. The vorticity
derivative in (\ref{r2}) has no electromagnetic analogue,
and vorticity appears to be an additional gravito-magnetic vector
potential. Furthermore, the gauge freedom in the
electromagnetic potentials does not have a direct gravitational
analogue in the Ricci gravito-potential equations
(\ref{r1}) and (\ref{r2}), 
since the gravito-electric/magnetic potentials
are invariantly defined kinematic quantities. (Note that the
Lanczos potential for the Weyl tensor does have a gauge
freedom analogous to that in the Maxwell four-potential \cite{dk}.)

The remaining Ricci equations in $1+3$ covariant form are \cite{mes}
\begin{eqnarray}
\dot{\Theta}+{\ts{1\over3}}\Theta^2 &=& -{\ts{1\over2}}(\rho+3p)
+\D^a\dot{u}_a+\dot{u}^a\dot{u}_a+2\omega^a\omega_a-\sigma^{ab}
\sigma_{ab} \,, \label{r3}\\
\dot{\omega}_{\langle a\rangle}+{\ts{2\over3}}\Theta\omega_a &=&
-{\ts{1\over2}}\c\dot{u}_a+\sigma_{ab}\omega^b \,,\label{r4}\\
{\ts{2\over3}}\D_a\Theta &=& -\c\omega_a+\D^b\sigma_{ab}
+2[\omega,\dot{u}]_a \,, \label{r5}\\
\D^a\omega_a &=& \dot{u}^a\omega_a \,, \label{r6}
\end{eqnarray}
and do not involve the gravito-electromagnetic field.

Finally in this section, we extend the analogy to wave propagation.
The magnetic wave equation (\ref{wave1}) 
has a simple gravito-magnetic analogue. In order to isolate the
purely tensor perturbations of an FLRW background in a
covariant (and gauge-invariant) way, one imposes $\omega_a=0$
\cite{he}. We linearize and take the curl of equation
(\ref{eq:edot}), using the
linearizations of equations (\ref{eq:hdot}) and 
(\ref{eq:divh}), and identities
(\ref{id7}) and (\ref{id11}). This
does not directly produce a wave equation, since the curl
of the shear
term in (\ref{eq:edot}) has to be eliminated. (In the Maxwell case
this feature did not arise, since we set $j_a=0$.)
The elimination is achieved via the Ricci equation (\ref{r2}), and
we find that
\begin{equation}
\Box^2H_{ab}\equiv-\ddot{H}_{ab}+\D^2H_{ab} \approx
7H\dot{H}_{ab}+2\left(3H^2-p\right)H_{ab}\,,
\label{wave2}\end{equation}
in agreement with \cite{he,DBE96}, and in striking analogy with
the magnetic wave equation (\ref{wave1}). 
Further discussion of 
covariant gravitational wave theory may be found in
\cite{he,DBE96,mes,bb}.

\section{Gravitational monopoles}

The electromagnetic correspondence we have developed
suggests a covariant
characterization of gravito-electric (magnetic) {\em monopoles}, 
as stationary vacuum spacetimes outside isolated sources,
with purely electric (magnetic) free gravitational field, i.e.,
$H_{ab}=0$ ($E_{ab}=0$). This is reinforced by 
the fact that monopoles do not radiate, and
gravitational radiation necessarily
involves both $E_{ab}$ and $H_{ab}$ nonzero (see 
\cite{EH96,DBE96,mes}, consistent with Bel's criterion 
$P_a\neq0$ \cite{bel,zak}). 
Our identification in the previous section  
of density inhomogeneity and angular momentum
density as sources of, respectively, gravito-electric and
gravito-magnetic fields, suggests that the monopole
sources will be respectively mass and angular momentum. However,
as pointed out previously, it is possible that non-zero
angular momentum is compatible with a purely
gravito-electric field, as illustrated by the G\"odel solution.

The four-velocity field $u^a$ is not defined by a fluid, but is
defined as the normalization of the stationary Killing
vector field $\xi^a=\xi u^a$. 
As a consequence of Killing's equations, 
we have $\Theta=0=\sigma_{ab}$
\cite{kramer}, so that
\[
\nabla_bu_a=\ep_{abc}\omega^c-\dot{u}_au_b \,.
\]
The covariant equations governing non-perturbative
monopoles are complicated. Some simplification
arises from the Killing symmetry, which implies
\begin{eqnarray*}
{\cal L}_\xi\omega_a &=&\xi\dot{\omega}_a+u_a\omega^b\D_b\xi=0\,, \\
{\cal L}_\xi H_{ab} &=& \xi\dot{H}_{ab}+2\xi\omega^c\ep_{cd(a}
H_{b)}{}^d-2\xi u_{(a}H_{b)c}\dot{u}^c=0 \,,
\end{eqnarray*}
and a similar equation for $E_{ab}$. Then it follows that
\begin{eqnarray}
\dot{\omega}_{\langle a\rangle} &=& 0 \,,\label{l1} \\
\dot{H}_{\langle ab\rangle} &=& -2\omega^c\ep_{cd(a}H_{b)}{}^d
\,, \label{l2} \\
\dot{E}_{\langle ab\rangle} &=& -2\omega^c\ep_{cd(a}E_{b)}{}^d
 \,. \label{l3}
 \end{eqnarray}

Now equations (\ref{l1})--(\ref{l3}), together with the basic
monopole conditions, are applied to the Bianchi equations
(\ref{eq:dive})--(\ref{eq:hdot}) and Ricci equations
(\ref{r1})--(\ref{r6}). We obtain:\\

{\em Gravito-electric and -magnetic monopoles:}
\begin{eqnarray}
\D^a\dot{u}_a &=& -\dot{u}^a\dot{u}_a-2
\omega^a\omega_a \,, \label{em7} \\
\dot{\omega}_{\langle a\rangle} &=& 0\,,\label{em7a}\\
 \c\dot{u}_a &=& 0
\,,\label{em8}\\
\c\omega_a &=& -2[\dot{u},\omega]_a \,,\label{em9} \\
\D^a\omega_a &=& \dot{u}^a\omega_a \,. \label{em10}
\end{eqnarray}

{\em Gravito-electric monopole:}
\begin{eqnarray}
\D^bE_{ab} &=& 0 \,, \label{em1} \\
0 &=& E_{ab}\omega^b \,, \label{em2} \\
\dot{E}_{\langle ab\rangle} &=& 0\,,\label{em2a}\\
0 &=& \omega^c\ep_{cd(a}E_{b)}{}^d
\,, \label{em3} \\
\c E_{ab} &=& -2\dot{u}^c\ep_{cd(a}E_{b)}{}^d \,, \label{em4}\\
E_{ab}-\D_{\langle a}\dot{u}_{b\rangle} &=&
 \dot{u}_{\langle a}\dot{u}_{b\rangle}
 -\omega_{\langle a}\omega_{b\rangle} \,, \label{em5}\\
 \D_{\langle a}\omega_{b\rangle} &=&
  -2\dot{u}_{\langle a}\omega_{b\rangle} \,. \label{em6}
\end{eqnarray}

{\em Gravito-magnetic monopole:}
\begin{eqnarray}
\D^bH_{ab} &=& 0 \,, \label{mm1} \\
0 &=& H_{ab}\omega^b \,, \label{mm2} \\
\dot{H}_{\langle ab\rangle} &=& 0\,,\label{mm2a}\\
0 &=& \omega^c\ep_{cd(a}H_{b)}{}^d
\,, \label{mm3} \\
\c H_{ab} &=& -2\dot{u}^c\ep_{cd(a}H_{b)}{}^d \,, \label{mm4}\\
\D_{\langle a}\dot{u}_{b\rangle} &=&
- \dot{u}_{\langle a}\dot{u}_{b\rangle}
 +\omega_{\langle a}\omega_{b\rangle} \,, \label{mm5}\\
H_{ab}- \D_{\langle a}\omega_{b\rangle} &=&
  2\dot{u}_{\langle a}\omega_{b\rangle} \,. \label{mm6}
\end{eqnarray}

Equation (\ref{em8}) implies that there exists
an acceleration potential:
\begin{equation}
\dot{u}_a =\D_a\Phi \,.
\label{ap}\end{equation}
This holds even when $\omega_a\neq 0$, despite the identity
(\ref{id1}), since $\Phi$ is invariant under $\xi^a$, so that
$\dot{\Phi}=0$. Equation (\ref{em9}) shows that $\c\omega_a$
is orthogonal to the vorticity and four-acceleration:
\[
\omega^a\c\omega_a=0=\dot{u}^a\c\omega_a \,.
\]

Schwarzschild spacetime, where also $\omega_a=0$
(since staticity implies $u^a$ is hyper-surface orthogonal),
is clearly a  non-perturbative gravito-electric monopole
according to our covariant definition: it 
is a static vacuum spacetime satisfying $H_{ab}=0$, by 
virtue of the Ricci equation (\ref{r2}).
Equations (\ref{ap}) and (\ref{em7}) imply
\begin{equation}
\D^2\Phi+\D^a\Phi\D_a\Phi =0\,.
\label{phi}\end{equation}
The solution $\Phi$ determines $\dot{u}_a$ and $E_{ab}$, and
equation (\ref{phi}) ensures that the monopole conditions
(\ref{em7})--(\ref{em6}) are identically satisfied.
 
It is not clear whether there exist 
consistent non-perturbative gravito-magnetic monopoles,
i.e. spacetimes satisfying the covariant equations
(\ref{em7})--(\ref{em10}) 
and (\ref{mm1})--(\ref{mm6}).\footnote{In
\cite{mawh} it is shown that non-flat
vacuum solutions with purely magnetic Weyl tensor are a very
restricted class, and it is suggested that there may be no
such solutions.}
However, linearized
gravito-magnetic monopoles have been found, for example the
Demianski-Newman solution \cite{dn} (see below). It is
also not clear whether there exist gravito-electric
monopoles with angular momentum (i.e. $\omega_a\neq 0$).

In the case of linearization about a
flat Minkowski
spacetime, the right-hand sides of equations (\ref{em7})--(\ref{mm6})
may all be set to zero. In particular,
equation (\ref{em9}) implies that there is a vorticity potential:
\begin{equation}
\omega_a \approx \D_a\Psi \,.
\label{vp}\end{equation}
The linearization of equations (\ref{em5}) and (\ref{mm6}),
together with the scalar potential equations (\ref{ap}) and
(\ref{vp}), then imply that the curls vanish to linear order.
Thus the linearized gravito-electric monopole
is covariantly characterized by equations (\ref{ap}),
(\ref{vp}) and
\begin{eqnarray}
\D^2\Phi\approx 0\,,~~E_{ab}\approx \D_a\D_b\Phi\,,
~~ \D_{\langle a}\D_{b\rangle}\Psi\approx 0\,, \label{le2}
\end{eqnarray}
while for the linearized gravito-magnetic monopole
\begin{eqnarray}
\D^2\Psi\approx 0\,,~~H_{ab}\approx\D_a\D_b\Psi\,, ~~
\D_{\langle a}\D_{b\rangle}\Phi\approx 0 \,. \label{lm2}
\end{eqnarray}

It follows in particular that a linearized 
non-rotating gravito-electric monopole is mapped
to a linearized non-accelerating gravito-magnetic monopole via
\begin{equation}
\i_{ab}\rightarrow i\,\i_{ab}\,,~~\omega_a \rightarrow \dot{u}_a\,,~~
\dot{u}_a\rightarrow -\omega_a \,.
\label{map}\end{equation}
Linearized Schwarzschild spacetime
is readily seen to satisfy equation (\ref{le2})
with
$\Phi=-M/r$, where $M$ is the mass and $r$ the area coordinate. 
Using the spatial duality rotation and kinematic interchange described
by equation (\ref{map}), this monopole is mapped to
a linearized non-accelerating gravito-magnetic monopole with 
potential $\Psi=-M/r$.
In comoving stationary coordinates, the metric of the
linearized magnetic monopole follows from $\dot{u}_a=0$ and
$\omega_a=\D_a\Psi$, using a theorem in \cite{ell73} (p 24):
\begin{eqnarray}
ds^2&=&-dt^2+dr^2+r^2\left(d\theta^2+\sin^2\theta\, d\varphi^2\right)
 + 4M \cos\theta\, d\varphi\, dt\,.
\label{eq:tn2}
\end{eqnarray}
This is a Taub-NUT solution with $m = 0$, $\ell = 
-M$ and linearized in $\ell$ (\cite{kramer}, p.133; see
also \cite{dowker}).
In fact, this is precisely the linearized solution found in 
\cite{dn}, so that we have a covariant characterization of that 
solution in the framework of gravitational duality.
Clearly the magnetic `charge' 
$M$ is an angular momentum parameter, not a mass parameter, 
and the metric  in
equation (\ref{eq:tn2}) describes an isolated source with
angular momentum but no mass.

\section{Concluding remarks}

A covariant $1+3$ approach, based on \cite{ell71} and its
extension \cite{roy,mes}, is ideally suited to an analysis
of the free gravitational field that is based on 
observable physical and geometric quantities, with a clear
and transparent analogy in well-established electromagnetic
theory. We have used such an approach, including in particular
the generalization of covariant spatial vector analysis to
spatial tensor analysis, which involves developing a
consistent covariant definition of the tensor curl and its
properties. Via this approach, we showed the remarkably
close analogy between the Maxwell equations for the electric/magnetic
fields and the Bianchi identities for the gravito-electric/magnetic
fields. Although this analogy has long been known in general terms,
our approach reveals its properties at a
physically transparent level, with a detailed accounting for
each physical and geometric quantity.
We found new interpretations of the role
of the kinematic quantities -- expansion, acceleration, vorticity
and shear -- in the source and coupling terms of
gravito-electromagnetism. The tracefree part of the Ricci identities
also reveals the role of the kinematic quantities as
gravito-electric/magnetic potentials.

The analogy provides a simple interpretation of the
super-energy density and super-Poynting vector as natural
$U(1)$ invariants, and we derived the exact nonlinear
conservation equation that governs these quantities,
and which involves a further natural invariant, i.e. the anisotropic
super-pressure.
We also used the analogy to show that a covariant spatial
duality invariance exists in vacuum gravito-electromagnetism,
precisely as in source-free electromagnetism. Duality invariance
has been important in some recent developments in field and string
theory, and the gravito-electromagnetic invariance in the form
found here may also facilitate new insights into gravity.
A crucial feature in the gravitational case, arising from its
intrinsic nonlinearity, is that the duality invariance does
not map Einstein solutions to Einstein solutions, since
the Ricci identities are not invariant. Further work is needed
to investigate whether a simultaneous geometric or kinematic
transformation can be found, so that the Bianchi and Ricci
equations are invariant under the combined transformation.

We showed that in linearized vacuum gravity, there is
a simple combined duality/ kinematic transformation that
maps the Schwarzschild gravito-electric monopole
 to the Demianski-Newman gravito-magnetic monopole. This
 covariant characterization of the relation between these
 linearized solutions was based on our covariant definition
 of gravito-monopoles in the general nonlinear theory. Further work
 is needed on the governing equations for these monopoles,
 in particular to see whether nonlinear gravito-magnetic
 monopole solutions may be found. A better understanding of
 the relation between nonlinear
 gravito-electric/magnetic monopoles could, as in field
 theory, open up new approaches and insights.

\newpage 

\ack
We thank Bill Bonnor,
Marco Bruni, George Ellis, Stuart Dowker, Stefano Liberati,
Donald Lynden-Bell, 
David Matravers, Dennis Sciama and Claudio  Scrucca for 
very helpful discussions and comments.

\end{document}